# Data-Driven Model for Failure Analysis of Internet of Things Devices : A Preliminary Study

Thanitnan Kammuang
Department of Control System and
Instrumentation Engineering
King Mongkut's University of
Technology Thonburi (KMUTT)
Bangkok, Thailand

Watthanai Luealamai
Department of Control System and
Instrumentation Engineering
King Mongkut's University of
Technology Thonburi (KMUTT)
Bangkok, Thailand

Issarapong Khuankrue
Department of Control System and
Instrumentation Engineering
King Mongkut's University of
Technology Thonburi (KMUTT)
Bangkok, Thailand

*Abstract*— This paper proposes the preliminary study of the data-driven failure analysis model on the internet of things (IoT) devices. This model focus on the impact of data transferring both get and receiving data on class C of Low Power Wide Area Network (LoRaWAN). To set up the network, the authors develop the combination of four several technology parts, including 1) the End Device Gateway Network server of LoRa IoT, 2) an Application server for store the data into the database, 3) the Dashboard to show and got the command by the user, and 4) the failure analysis model based on Bayesian belief networks which calculate the probability values that collect the data transferring both uplink and downlink on the network connection in this study. In the testing phase, the authors input the separated data into the data-driven failure analysis model to analyze the time and latency of the connection by the concern of the impact and risk for the failure of the overall system. The model will show the probability value of failure. The authors hope to use the results to clarify that a tested IoT device suitable for use or not.

## I. Introduction

In the industry 4.0 policy of Thailand, the new era of technology is the core concept to develop the country. The control system, automation, and industrial informatics are the main fields which require to transform many industry sectors. The internet of things devices and wireless sensor networks have been increasing the amount of attention both in industrial and household areas. The core technology used to design the control systems based on various wireless connection standards such as Wireless HART, ISA100 Wireless Network Technology, ZigBee Wireless Network Technology, and Long Range Wireless Network (LoRaWAN). The LoRaWAN is a wireless technology which us a cloud-based medium access control data rate and power that be able to apply to advanced instrumentation and control equipment. Because of the simplicity in installation and maintenance, the low energy consumption, and has high reliability by calling these things as a whole Internet of Things (IoT). However, the major problem of using these devices is the delay time between the devices and the server that used for handling and storage.

On the concept of technical debt [1], in software development, the cost will increase because of additional rework by the wrong decision making or decide by a lack of data support. In case of IoT devices, it was chosen instead of using a better approach, especially the network latency, that would take longer, will be the technical debt at last. The duration of data transmission is latency. In industrial plants, the performance of time has required the accuracy and speed of use of equipment. The technology being adopted requires a fast transmission time to improve their production performance by increase the number of IoT or wireless devices into their traditional production. Moreover in the factory, there may be waves that interfere with the operation of wireless devices and make the use of the wireless sensor network more likely to malfunction, such as incorrect data transmission. Data has been missing or no data transmission occurs. In fact, the device transmits with high latency, causing it to slow down or cause internal failure. Therefore, the data transmission among the wireless network devices, the server, and the probability analysis of the delay time as a determinant for the acceptance.

The machine learning model can apply to model the decision support system for the IoT devices in the factory. Siryani and others proposed the framework that operates on the IoT environment. [3] This framework improves the cost prediction for the meter field operation and presents the recommendation base in the electric smart meter issue. They make a comparison on the Bayesian network with the Naïve Bayes, random forest, and decision tree. In the same way, on hydroponics smart farming, Alipio and others also apply the Bayesian network into the inference system on the connection of IoT devices. They make an experiment that shows it will improve manual control which implies that the use of exact inference in Bayesian network can support in producing high-quality crops.[4]

This paper proposed the preliminary study of the data-driven failure analysis model on IoT devices. The technology adopted the probability of occurrence suitable for application in that industry to be the failure analysis is to bring the transmission data. This model analyses and looking for the results and decision-making. The data was collected and inked its insight into the failure of the system. The Bayesian Network is a model used to describe cause-effect relationships through the use of probability theory and expressed in the form of a cycle graph, direction, and condition table. It can figure out the probability of a transmission failure. This model is a first step to design the data-driven model for the judgment of any IoT devices which can use at any time depending on the data that can be collected. It will be impacted the changing of technology or device that can better respond to the industry or area-based measurements.

This study expects to understand the failure analysis techniques used to analyze the Internet of Things devices. It will be able to use to create models for failure prediction. The model is the data-driven decision on any equipment or machinery that needs to embed the IoT into its. In the industry side, it can be reduced the cost and time for maintenance and other operations of industrial systems.

## II. MODELLING THE DATA-DRIVEN FAILURE ANALYSIS

### A. Setting up

To set up the network, the authors develop the combination of four several technology parts, including 1) the End Device Gateway Network server 2) an Application server for store the data into the database, 3) the Dashboard to show and got the command by the user, and 4) the Failure analysis model which calculate the probability values that collect the data transferring both uplink and downlink on the network connection in this study. The details of each part are shown as follows: -

- **LoRa IoT** stands for the Long Range, which derives from the need for wide-range, low-power (LPWAN) wireless communications. It refers to the connection protocol only on the part of the connection. So, LoRaWAN is the way to interconnection in a network with a wide communication distance by using the unlicensed frequencies and reduce energy consumption, which compared to Wi-Fi and Bluetooth. Therefore, private LoRa networks are ideal for connecting smart sensors and other low energy devices and transmitting data over a long distance. This study sets up the end device by using the LoRaWan gateway network server.

- **MongoDB** is the application and database server used to store the accessed data with the user ID to identify the processing of each user. MongoDB is a NoSQL database, which is storing data in JSON format. This strength of the database is the speed and be storable the value as key and value.

- **Grafana** is a dashboard service, which displays the graphs of various data. it extracts the data in real-time, where it can retrieve the data from a very popular data source.

- **Basic Bayesian networks** is a graphical modal that uses the principles of probability. It is expressed in two parts: 1) The Conditional Probability Table (CPT) and 2) Directed Acyclic Graph (DAG) by the structure of the graph. This is comprised of a set of nodes and connections, each of which is represented by a random variable. While the connection between each node represents the dependency of random variables. The conditional autonomy in a graph is estimated using statistical, mathematics, and computer knowledge. This study uses the basic Bayesian belief networks as the failure analysis model.

### B. Data Transfer and Collection

This study aims to analyze the functionality of a wireless connection system for the Internet of Things (IoT). IoT is the application to transmit and receive data via LoRa technology. It has to transmit the data at a distance of 5-15 kilometers. The overall operation of the system divided the system into 2 main parts: 1)The hardware part consists of a structure, a storage device, and an electronic device. 2) The software part consists of the device data collection, the data transmission via LoRa technology, the database storage, the performance display ,and the data analysis. As shown in Fig 1, the data transmission of data on the devices will send data to the network server and pass data from the network server to the application server, which stored on the database. To show results, the user can see that various functions on the website, requiring an intermediary to pass data from the website to the main program using JavaScript language, which developed on Node.js with other application software.

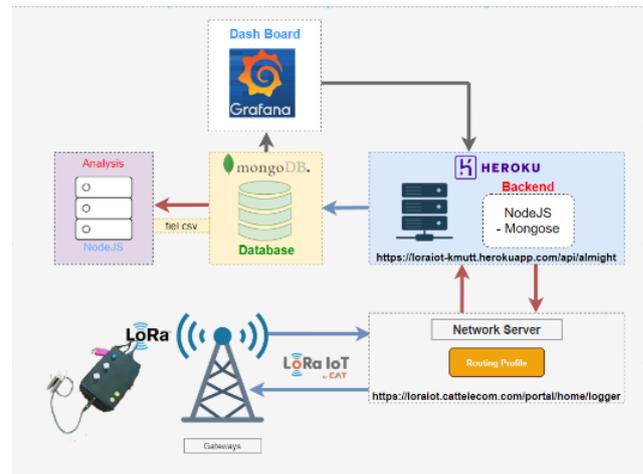

Fig. 1. Overview of the data transfer and collection

### C. Proposed Failure Analysis Model

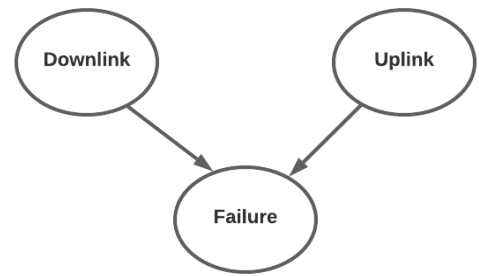

Fig. 2. Proposed failure analysis model using baisc bayesian networks

After receiving data from MongoDB and exporting the file to the comma-separated value (CSV) file, it will show the required Timestamp information, there is also additional information. In creating a good model, there should be data cleaning, which is the process of reviewing and editing (or removing) unnecessary items. In this model, only the timestamp on the device and server is used. The data is divided into 2 parts for analysis:

- **Uplink Data** obtained from uplink by operating the device microcontroller for the sensor working and perform the measurement from the sensor and forward it to the server.

- **Downlink Data** obtained from downlink by ordering through the dashboard to send control commands to the LED device attached to run on the microcontroller devices (IoT).

The timestamp between device and server is the subtract the time between the server and the device each time of the data transmission, calculated in terms of time in seconds. Then, the calculation of the probability of time exceeding the acceptance condition. The probability of time exceeding the accepted condition (P(T)) is the proportion of time greater than the set condition time to the total amount of information, which must take the data set between the server and device and loop to calculate that each data.

To analyze the failures that occur, the basic Bayesian networks model was modified as Fig 2. This model will consist of independent factors or random variables. The probability of the time that exceeds the accepted conditions, including uplink node, downlink node, and failure. All random variables, counted as nodes, form a network through the use of connections and each random variable has a table that distributes the probabilities of the conditions in that variable. The probability of each value arises from data collection. Due to the data collection constraint, the study can only collect data of the Internet's transmission time beyond the overtime condition, which allows determining the probability of the time, which is exceeded only the acceptable conditions. Therefore, the probability of all can only be obtained from randomization and appropriate testing. On Bayes's theorem, it will find a solution through a conditional value, assuming nodes Y stand for uplink and Z stand for downlink, which are parent nodes of node X or failure. Then, a conditional equation can be established. As follows the equation (1), is used to determine the probability of events represented by child nodes arising from an event represented by parent nodes that may or may not specify the condition of the event.[4][5]

$$P(Y, Z) = \sum_{X,Y,Z} \frac{P(X,Y,Z)}{P(X,Y)} \qquad (1)$$

*D. Preliminary Test and Result*

The preliminary test results, as shown in Table I to III, shown that the data acquisition software is compliant with the predetermined development scope. The timestamp on the device and server is used:1) Uplink Data are 87 records, and 2) Downlink data are 782 records. On the uplink latency node, the data set calculates the probability through the specified time condition set at the greater than or equal to 37 seconds when the data was split in half, 50 % for the Training set and the other half for the Test set. The result of probability on Training set: P (U = 1) = 0.503, or the probability of uplink latency greater than or equal to 37 seconds is equal to 0.503. The result of probability on Test set: P (U = 1) = 0.38 or the probability of uplink latency greater than or equal to 37 seconds is equal to 0.38 as shown on Table I.

In the same way on the downlink latency node, the data set calculates the probability through the specified time condition set at the greater than or equal to 42seconds when the data was split in half, 50 % for the Training set and the other half for the Test set. The result of probability on Training set: P (D = 1) = 0.605, or the probability of downlink latency greater than or equal to 42 seconds is equal to 0.605. The result of probability on Test set: P (D = 1) = 0.791 or the probability of downlink latency greater than or equal to 42 seconds is equal to 0.791 as shown on Table II. Functionality can be tested through the handling of data obtained from CAT's Network Server (CAT TELECOM) and displayed in the MongoDB database and display on the Grafana platform.

For the use of the storage device, it is necessary to upload the recording time to the new device every time before use. The stability of receiving commands from the monitor is still not stable enough. Graph display in the monitor screen needs to be reset all the time in order to retrieve the values to display on the dashboard. So it can be concluded that This data collection tool is still very unstable. The accurate in receiving-delivery orders of data collection because the recording time does not correspond to the real-time of about 60 minutes, including the remote connection system of LoRa IoT by CAT, and sometimes the token request for transmission does not go through making the data unable to be exported.

In Table III, the way to find the accuracy of this model, the authors use the confusion matrix as the accurate analysis by dividing the data set into a training dataset and a test dataset. It will be able to find the confusion matrix by assigning the training dataset to predicted data and the test dataset as actual data. The results are True Positive (TP) = 0.7485, True Negative (TN) =0.982, False Positive (FP) = 0.018, and False Negative (FN) =0.2515. While the accuracy = (TP + TN) / (TP+TN+FP+FN), the accuracy is 0.8652.

TABLE I.  TRAIN DATASET RESULT

| Uplink Latency Node | |
|---|---|
| P(Uplink = 1) | 0.503 |
| P(Uplink = 0) | 0.497 |
| **Downlink Latency Node** | |
| P(Downlink = 1) | 0.605 |
| P(Downlink = 0) | 0.395 |

TABLE II.  TEST DATASET RESULT

| Uplink Latency Node | |
|---|---|
| P(Uplink = 1) | 0.380 |
| P(Uplink = 0) | 0.620 |
| **Downlink Latency Node** | |
| P(Downlink = 1) | 0.791 |
| P(Downlink = 0) | 0.209 |

TABLE III.  FAULURE ANALYSIS RESULTS

| Failure Node | | | |
|---|---|---|---|
| Uplink | Down-link | P(Failure =1|Uplink,Downlink) | P(Failure =0|Uplink,Downlink) |
| 1 | 1 | 0.605 | 0.395 |
| 1 | 0 | 0.093 | 0.907 |
| 0 | 1 | 0.267 | 0.733 |
| 0 | 0 | 0.023 | 0.977 |

III. CONCLUSION

In conclusion, this paper proposes the preliminary study of the data-driven failure analysis model on the internet of things (IoT) devices. A system is a setup between a wireless device using LoRaWAN technology to connect to the Gateway of CAT and another server, which can transfer the data to a simulation server built through the Heroku platform and stored at MongoDB. It was found that the device could send and receive commands from the Server, that is, the Uplink data set and the Downlink data set only in some areas. The signal without the LoRaWAN network signal makes it complex to collect the data, so simulated latency is used for analysis. In terms of the results of the simulated latency data of the two data sets, it was found that the model was able to provide an accuracy of 86.52%. Note that there are fewer data sets that can be modified if there is more time to collect data.